# Thickness Dependent Magnetic Transition in Few Layer 1T Phase CrTe$_2$

Pengfei Gao,[†] Xingxing Li,[*,†,‡] and Jinlong Yang[*,†,‡]

[†]Synergetic Innovation Center of Quantum Information and Quantum Physics, University of Science and Technology of China, Hefei, Anhui 230026, China

[‡]Hefei National Laboratory for Physical Sciences at the Microscale, University of Science and Technology of China, Hefei, Anhui 230026, China

*Supporting Information Placeholder*

**ABSTRACT:** Room temperature two-dimensional (2D) ferromagnetism is highly desired in practical spintronics applications. Recently, 1T phase CrTe$_2$ (1T-CrTe$_2$) nanosheets with five and thicker layers have been successfully synthesized, which all exhibit the properties of ferromagnetic (FM) metals with Curie temperatures around 305 K. However, whether the ferromagnetism therein can be maintained when continuously reducing the nanosheet's thickness to monolayer limit remains unknown. Here, through first-principles calculations, we explore the evolution of magnetic properties of 1 to 6 layers CrTe$_2$ nanosheets and several interesting points are found: First, unexpectedly, monolayer CrTe$_2$ prefers a zigzag antiferromagnetic (AFM) state with its energy much lower than that of FM state. Second, in 2 to 4 layers CrTe$_2$, both the intralayer and interlayer magnetic coupling are AFM. Last, when the number of layers is equal to or greater than five, the intralayer and interlayer magnetic coupling become FM. Theoretical analysis reveals that the in-plane lattice contraction of few layer CrTe$_2$ compared to bulk is the main factor producing intralayer AFM-FM magnetic transition. At the same time, as long as the intralayer coupling gets FM, the interlayer coupling will concomitantly switch from AFM to FM. Such highly thickness dependent magnetism provides a new perspective to control the magnetic properties of 2D materials.

In recent years, spintronics, which uses electron spin to transmit and process information, has developed vigorously due to its great advantages of fast data processing and low energy consumption.[1-4] Two-dimensional (2D) materials with room temperature ferromagnetism are crucial in developing practical nano-spintronic devices, which however remain a big challenge. Up to now, only a few 2D materials have been confirmed experimentally to be ferromagnetic around 300 K, such as 2D films of VTe$_2$[5] and VSe$_2$.[6-7] Through electrical gating, the ferromagnetic Curie temperature $T_c$ of 2D Fe$_3$GeTe$_2$ can be enhanced to room temperature.[8] Few-layer Fe$_5$GeTe$_2$ has also been synthesized,[9] with a $T_c$ of about 270 K, slightly lower than room temperature. Theoretically, although several room temperature ferromagnets have been predicted, such as organometallic ferrimagnetic semiconductors,[10-11] transition metal embedded borophene nanosheets,[12] 2D Fe$_2$Si nanosheet[13] and alloy compounds,[14] their experimental realization is still hard.

Recently, Sun et al. have successfully prepared ultra-thin two-dimensional 1T-CrTe$_2$ nanosheets by mechanical exfoliation. Experimental measurements revealed that few layer 1T-CrTe$_2$ are ferromagnetic (FM) metals whose ferromagnetism can hold above 300 K,[15] thus they are promising candidates for spin injectors working at room temperature. However, the experimentally prepared ultra-thin CrTe$_2$ has at least 5 layers, and whether the ferromagnetism of CrTe$_2$ with a further reduced thickness can be maintained remains an interesting question, and needs to be figured out for better understanding and application of this material.[16-17]

For this purpose, based on first principles calculations, we here systematically study the magnetic ground states of 1T-CrTe$_2$ with number of layers varying from 1 to 6, and the magnetism is found to be highly dependent on the nanosheet's thickness: an intralayer and interlayer AFM-FM magnetic transition are revealed and occur at a critical thickness of five CrTe$_2$ layers. Therefore, CrTe$_2$ nanosheets with 1 to 4 layers are all AFM metals without spin polarization. The bulk's FM metal property only starts to emerge at five layers CrTe$_2$ nanosheets.

As the starting point, we study the electronic and magnetic properties of bulk CrTe$_2$. The detailed computational methods can be found in the supporting information. A series of van der Waals (vdW) correction methods are tested, among which the opt-B86b-vdW method[18-19] describes the crystal structure of the bulk phase best (See Table S1 for details). Therefore, the results based on the opt-B86b-vdW functional are reported. The bulk CrTe$_2$ [Figures 1(a) and (b)] belongs to the hexagonal crystal system with optimized lattice parameters a = 3.789 Å and c = 6.079 Å, which are very close to the experimental values (a = 3.789 Å and c = 6.096 Å).[20] Cr atoms with +4 formal charges are located in the octahedral crystal field formed by Te atoms. The CrTe$_2$ layers are packed along z direction in an AA pattern with an interlayer distance 2.971 Å. Experimental measurement shows bulk 1T-CrTe$_2$ is a magnetic metal with a magnetic moment of 1.7 μB per formula and a Curie temperature of 310 K.[20] The calculated magnetic moment is 2.43 μB per formula, somewhat larger than the experimental value. When an effective Coulomb U value is incorporated, an even larger magnetic moment is obtained. Besides, when effective U value is greater than 1.5 eV, the ground interlayer magnetic coupling of bulk CrTe$_2$ is falsely predicted to be antiferromagnetic. Therefore, onsite U correction is not considered. The details are shown in Table S2. Using Monte Carlo method, we simulate the Curie temperature of bulk CrTe$_2$ based on the classical Heisenberg Hamiltonian model,

$$H = -\sum_{k}\sum_{i,j} J_k \times S_i \cdot S_j \quad (1)$$

where $J_k$ represent four magnetic exchange parameters, i.e. the nearest and next nearest neighbor exchange in CrTe$_2$ layer, and the nearest and next nearest neighbor exchange between CrTe$_2$ layers [Figures 1(a) and (b)]. The spin of Cr atom is approximately taken as S = 1. To obtain $T_c$, we calculate the specific heat $C_v = (\langle E^2\rangle - \langle E\rangle^2)/T^2$ at first after the system reaches equilibrium at a given temperature. Then $T_c$ is gained by locating the peak position in the $C_v$ plot. From the simulated $C_v$ curve in Figure 1(c), the Curie temperature is found to be 359 K, close to the experimental value of 310 K.

As a measurement of stability of spontaneous magnetization, the magnetic anisotropy energy (MAE) is also computed, which is defined as the total energy difference when magnetization is out-of-plane and in-plane: MAE = E[OUT] − E[IN]. The MAE without U is 0.7 meV/Cr, which means the easy magnetization axis of bulk CeTe$_2$ is aligned in the CrTe$_2$ plane, consistent with experimental observation. Figure 1(d) shows the density of states (DOS) of bulk CrTe$_2$, indicating that bulk 1T-CrTe$_2$ is a magnetic metal. Spin polarization around the Fermi level is predicted by the following equation:

$$\frac{\langle DOS_{up}(E_f) - DOS_{down}(E_f)\rangle}{\langle DOS_{up}(E_f) + DOS_{down}(E_f)\rangle} \times 100\% \quad (2)$$

and we obtain a spin polarization of 57%. The spin polarization of intrinsic CrTe$_2$ is somewhat low compared to that of ideal half metal (100%), and may be enhanced by external stimuli such as electrical gating or physiochemical modifications.

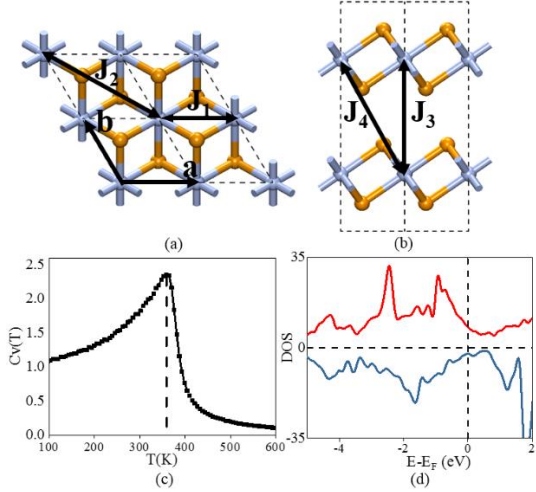

Figure 1. The structure of bulk 1T-CrTe$_2$ from (a) top and (b) side views. The blue-gray and dark yellow balls represent Cr and Te atoms, respectively. The four magnetic exchange interactions $J_k$ (k=1-4) are also indicated. (c) The simulated specific heat changing with temperature. (d) Density of states (DOS) of bulk 1T-CrTe$_2$. The red (blue) line represents the DOS of the up (down) spin channel. Fermi level is set to zero.

After resolving the properties of bulk 1T-CrTe$_2$ which can be viewed as a hypothetic nanosheet with infinity thickness, we then turn to the thinnest nanosheet, i.e. monolayer CrTe$_2$. In order to determine the magnetic ground state of monolayer CrTe$_2$, we construct a $2*2\sqrt{3}$ supercell and consider the following four magnetic states: intralayer ferromagnetic (FM), ABAB, AABB type striped antiferromagnetic (sAFM-ABAB, sAFM-AABB) and zigzag type antiferromagnetic (AFM-Z) configurations [Figures 2(a)-(d)]. Different from bulk CrTe$_2$, we find that the energies of the sAFM-ABAB, sAFM-AABB and AFM-Z states are 67.3 meV/Cr, 35.5 meV/Cr and 99.3 meV/Cr lower than that of FM state respectively, indicating that monolayer CrTe$_2$ is no longer FM. Instead, AFM-Z state is suggested to be the most stable among the four magnetic states we considered. Note that a recent experiment also revealed that monolayer CrTe$_2$ on graphene presents a zigzag type AFM order.[21] Previously, Adolfo et al. suggested a $\sqrt{3} \times \sqrt{3}$ reconstructed charge density wave (CDW) phase with trigonally distorted Cr atoms to be the ground state of monolayer CrTe$_2$.[17] However, we calculate the energy of this phase and find it is only 12.7 meV/Cr lower than that of FM state, and 86.6 meV/Cr higher than that of AFM-Z state. It should be pointed out that due to the limited number of magnetic states we considered, whether AFM-Z state is the global ground state for monolayer CrTe$_2$ can not be concluded at current stage. The density of states (DOS) for AFM-Z monolayer CrTe$_2$ is calculated and shown in Figure 2(e), from which one can see that monolayer CrTe$_2$ is an AFM metal with zero spin polarization in the vicinity of Fermi level.

By inspecting the structure of monolayer CrTe$_2$ (Table S3), we find that the in-plane lattice of AFM-Z state CrTe$_2$ (a' = 7.14 Å, b' = 12.30 Å) is about 6% contracted in both a' and b' directions compared with that of bulk phase (a' = 7.58 Å, b' = 13.13 Å). Such a big lattice variation is expected to produce significant modulation of the magnetic properties of CrTe$_2$. Therefore, in order to clarify the influence of lattice constant on the relative stability of different magnetic states, we calculate the energy variations of the four magnetic states, i.e. FM, sAFM-ABAB, sAFM-AABB and AFM-Z under different a' and b' lattices. Figure 2(f) shows the obtained phase diagram of most stable magnetic state changing with the lattices. The blue, gray and pink regions represent the sAFM-ABAB, AFM-Z and FM state is most stable, respectively. The lattice of pristine AFM-Z state CrTe$_2$ is represented by the black circle in Figure 2(f). It is clear that when the lattices in the a' and b' directions are both stretched properly, the intralayer magnetic coupling transits to FM. This is exactly the case for bulk CrTe$_2$, whose in-plane lattices in the a' and b' directions are both enlarged and located in FM region of the phase diagram [the white triangle indicated in Figure 2(f)].

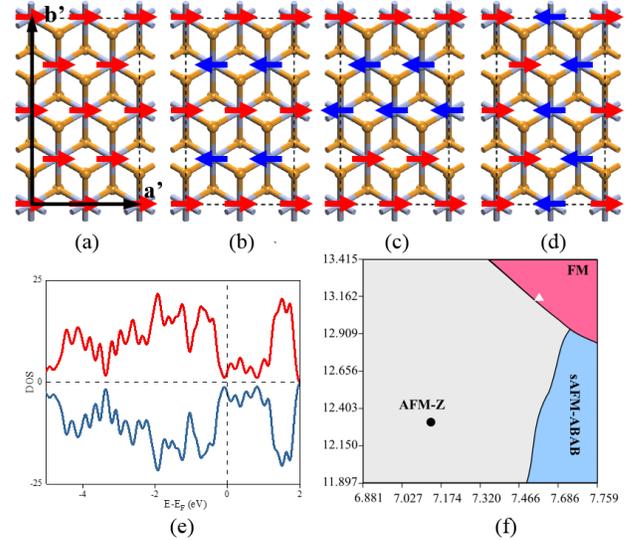

Figure 2. Top views of $2*2\sqrt{3}$ supercell with different intralayer magnetic orders, including (a) FM, (b) sAFM-ABAB, (c) sAFM-AABB and (d) AFM-Z. (e) DOS of AFM-Z state monolayer CrTe$_2$. The red (blue) line represents up (down) spin channel. Fermi level is set to zero. (f) Phase diagram of most stable magnetic state for monolayer CrTe$_2$ varied with lattice parameters in the a' and b' directions. The blue, gray and pink region represent sAFM-ABAB, AFM-Z and FM is the most energetically favorable, respectively. The black circle and white triangle represent the in-plane lattice location of monolayer and bulk CrTe$_2$, respectively.

Above results reveal that the ground magnetic coupling for bulk $CrTe_2$ is FM, while that for monolayer $CrTe_2$ is not. Therefore, how the magnetic state of $CrTe_2$ evolves from monolayer's AFM-Z to bulk's FM and what the critical number of layers for such a transition remain to be solved. Thus, few layer $CrTe_2$ nanosheets with 2 to 6 $CrTe_2$ layers stacked in bulk's AA pattern are systematically studied. To determine the magnetic ground state of few layer $CrTe_2$, eight different magnetic orders, i.e. four intralayer couplings (FM, sAFM-ABAB, sAFM-AABB, AFM-Z) combined with two interlayer couplings (FM, AFM) are considered. Taking the FM state as a reference, the calculated total energies under different magnetic states are shown in Figure 3. It can be seen that the most stable intralayer coupling of 2 to 4 layers $CrTe_2$ is still AFM-Z, similar to that of monolayer $CrTe_2$. When the number of layers is equal to or greater than five, the intralayer coupling becomes FM. Such an intralayer AFM-FM magnetic transition can be well explained by the phase diagram in Figure 2(f). In detail, the in-plane lattice of $CrTe_2$ increases with the number of layers, for example, the in-plane lattice of the most stable 4 layers $CrTe_2$ is a' = 7.22 Å, b' = 12.29 Å, which resides in the AFM-Z phase region, while the in-plane lattice of the most stable 5 layers $CrTe_2$ (a' = 7.54 Å, b' = 13.04 Å) changes to be at the border of the FM phase region. Further increasing the number of layers pushes the in-plane lattice approaching that of bulk, and they are all located in the FM phase region in Figure 2(f). Above all, the in-plane lattice variation is the main factor that causes the intralayer magnetic transition.

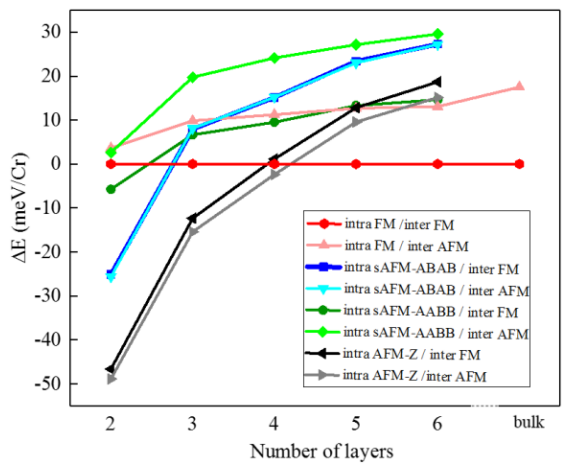

Figure 3. The total energies at eight different magnetic states by setting the energy of FM state to be zero. The FM, sAFM-ABAB, sAFM-AABB and AFM-Z before the slash indicate intralayer coupling, and the FM and AFM after the slash indicate interlayer coupling.

With regard to interlayer magnetic coupling, there also exists an AFM-FM transition: in 2 to 4 layers $CrTe_2$, an interlayer AFM coupling is preferred, which changes to be FM for 5 layers and thicker nanosheets (Figure 3). It is clear that the interlayer coupling of few layer $CrTe_2$ is related to the intralayer coupling directly: when the intralayer coupling is AFM-Z, the interlayer coupling would always prefer AFM. When the intralayer coupling gets FM, the interlayer coupling also transits to FM. It is worth mentioning that incorporating U correction (e.g. 1.2 eV) may alter the relative stabilities of eight different magnetic orders (Figures S1) and the critical number of layers corresponding to magnetic transition, but the fact of occurrence of intralayer and interlayer AFM-FM magnetic transition as the number of $CrTe_2$ layers increases remains unchanged.

Density of states (DOS) for few layer $CrTe_2$ are further calculated to investigate the evolution of electronic structures. As shown in Figure S2, it can be seen that the nanosheets with 2 to 4 $CrTe_2$ layers, which features both intra- and interlayer AFM coupling, all present no net electronic spin polarization (the same as monolayer $CrTe_2$), while the DOS of five layers and thicker $CrTe_2$ are quite similar to that of bulk $CrTe_2$, indicating that the FM metal property is unsensitive to the number of layers after surpassing the AFM-FM magnetic transition.

To summarize, based on first principles calculations, we uncover a highly thickness dependent magnetism in 2D 1T-$CrTe_2$ nanosheets. When the number of layers increases from 1 to 6, the $CrTe_2$ nanosheet undergoes an intralayer zigzag type AFM to FM transition, and an interlayer AFM to FM transition. Concomitantly, the electronic structure changes from an unpolarized metal to polarized metal. The lattice variation is responsible for such magnetic transitions. Our research indicates lattice strain and thickness can be potential routes to control and enrich the electronic and magnetic properties of 2D $CrTe_2$.

## ASSOCIATED CONTENT

### Supporting Information

Computational details; test of different vdW methods; test of different U values; The relative stabilities of eight magnetic orders with effective U of Cr being 1.2 eV; Density of states (DOS) for few layer $CrTe_2$.

## AUTHOR INFORMATION

### Corresponding Author

*E-mail: lixx@ustc.edu.cn
*E-mail: jlyang@ustc.edu.cn

### Notes

The authors declare no competing financial interest.

## ACKNOWLEDGMENT

This work is supported by the National Natural Science Foundation of China (Grant No. 21688102), by the National Key Research & Development Program of China (Grant No. 2016YFA0200604), by Anhui Initiative in Quantum Information Technologies (Grant No. AHY090400), by the Youth Innovation Promotion Association CAS (2019441) and by USTC Research Funds of the Double First-Class Initiative (YD2060002011). The computational resources are provided by the Supercomputing Center of University of Science and Technology of China, Supercomputing Center of Chinese Academy of Sciences, and Tianjin and Shanghai Supercomputer Centers.

Table of Contents Entry

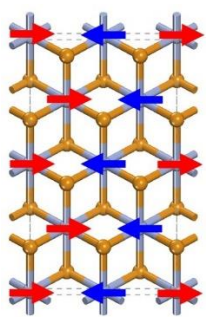 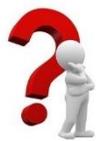 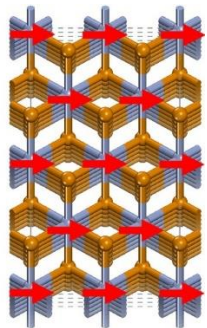